\documentclass[draft]{agujournal2019}
\usepackage{url} 
\usepackage{lineno}
\usepackage[inline]{trackchanges} 
\usepackage{soul}
\draftfalse

\journalname{Geophysical Research Letters}

\begin{document}

%
%


\title{Conservation laws for potential vorticity in a salty ocean or cloudy atmosphere}

%
%




\authors{Parvathi Kooloth\affil{1}, Leslie M. Smith\affil{1,2}, Samuel N. Stechmann\affil{1,3}}


\affiliation{1}{Department of Mathematics,}
\affiliation{2}{Department of Engineering Physics,}
\affiliation{3}{Department of Atmospheric and Oceanic Sciences, \\ University of Wisconsin--Madison, Madison, WI, 53706, USA}




\correspondingauthor{P. Kooloth}{parvathi.kooloth@wisc.edu}




\begin{keypoints}
\item Important conservation laws of circulation and potential vorticity are known only under idealized assumptions.
\item Here, conservation laws are described with additional
    realism, namely salinity in the ocean or moisture and clouds
    in the atmosphere.
\item Potential vorticity is not conserved for each fluid parcel but for certain pancake-shaped volumes.
\end{keypoints}

%
%

%
%


\begin{abstract}
One of the most important conservation laws in
atmospheric and oceanic science is conservation of
potential vorticity. The original derivation is
approximately a century old, in the work of
Rossby and Ertel,
and it is related to the celebrated circulation theorems of
Kelvin and Bjerknes.
However, the laws apply to idealized fluids,
and extensions to more realistic scenarios have been
problematic.
Here, these laws are extended to hold with additional
fundamental complexities, including salinity in the ocean,
or moisture and clouds in the atmosphere.
In the absence of these additional complexities,
it is known that
potential vorticity is conserved following each fluid parcel; here, for a salty ocean or cloudy atmosphere, 
the general conserved quantity is potential vorticity integrated over certain
pancake-shaped volumes.
Furthermore, the conservation laws are also 
related to a symmetry
in the Lagrangian, which brings a connection to
the symmetry-conservation relationships seen
in other areas of physics.
\end{abstract}

\section*{Plain Language Summary}
A vortex can be seen in the atmosphere or ocean in
a variety of settings, including the intense vortices
that are familiar from tornadoes and hurricanes.
A quantity called vorticity is a measure of the 
strength of a vortex, or, more generally, of the amount
of swirl in a fluid. Intimately connected to vorticity is an important conserved quantity referred to as potential vorticity. This quantity remains unchanged following the trajectory of a fluid parcel. Owing to its conserved nature, potential vorticity is widely used to study a wide range of weather events. However, this century-old conservation result is only valid in an idealized setting: an atmosphere devoid of clouds or an ocean without dissolved salt---i.e., without salinity. Here, the conservation law is extended to hold in a cloudy atmosphere and salty oceans. 
The generalized conservation statement presented here will aid in the wider applicability of potential vorticity for analyzing and understanding atmospheric and oceanic dynamics.
%
%

%


%
%
%
%

\section{Introduction}

\begin{figure}
    \centering
    \includegraphics[width=0.8\textwidth]{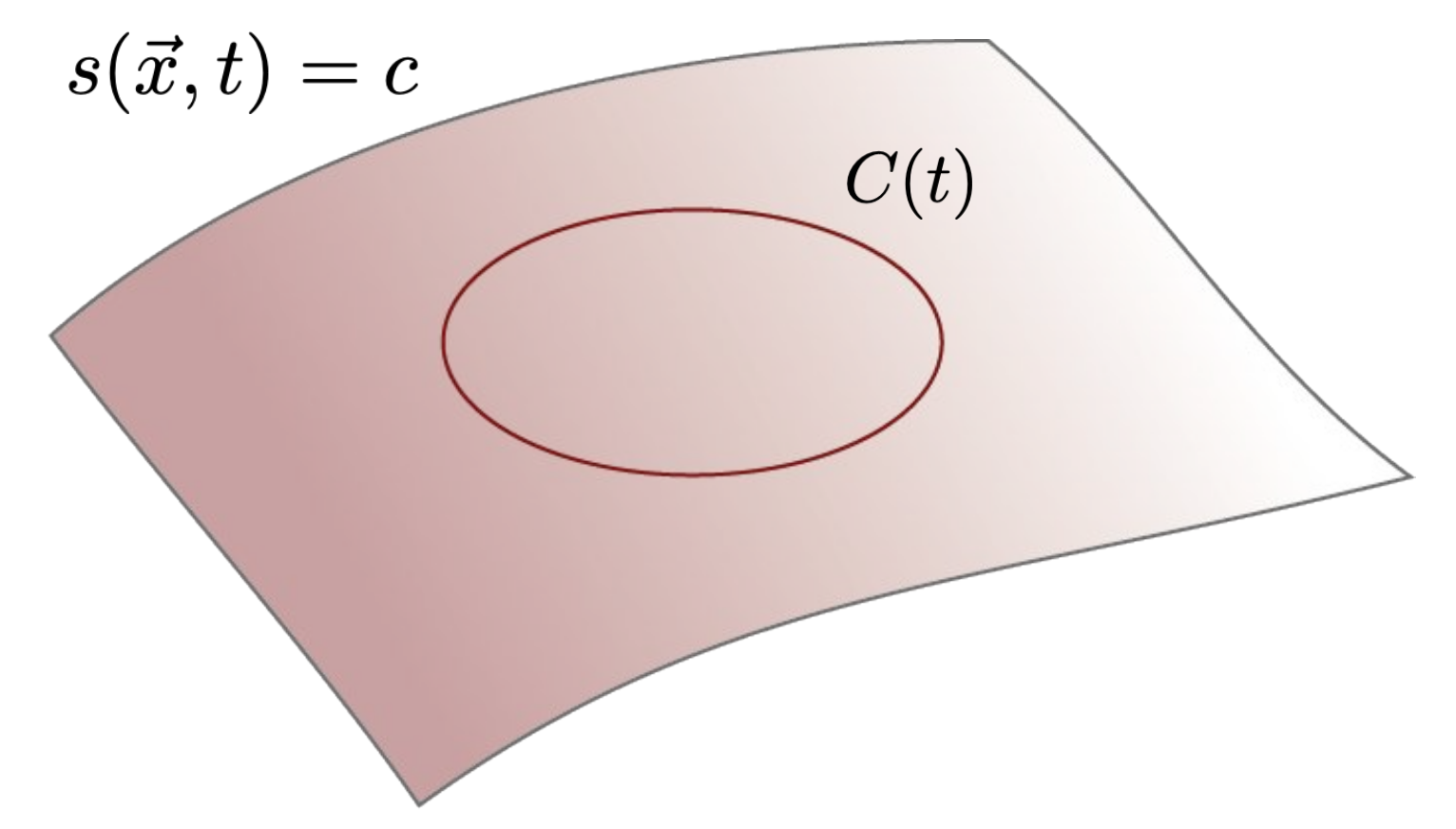}
    \caption{Geometry of the circulation theorem for a dry atmosphere or 
    freshwater ocean. The curve $C(t)$ may be any curve on a surface of constant entropy, 
    $s(\vec{x},t)=c$.}
    \label{fig:dry-circulation}
\end{figure}

In addition to conservation laws of mass, momentum,
and energy, other conservation laws arise for
fluids, such as conservation of circulation and 
potential vorticity.
The circulation $\Gamma$ is defined as the integral of the
velocity $\vec{u}$ along a closed curve $C$,
\begin{linenomath*}
\begin{equation}
    \Gamma=\oint_{C} \vec{u}\cdot d\vec{l},
    \label{eqn:circ-def}
\end{equation}
\end{linenomath*}
and the circulation theorems of 
Kelvin and Bjerknes \cite{thomson1867,bjerknes1898hydrodynamischen,thorpe2003bjerknes} show that, under certain assumptions,
the circulation is conserved if the closed curve $C(t)$
moves with the fluid. 
See Figure~\ref{fig:dry-circulation} for an illustration.
The circulation is also related to
the vorticity, $\vec{\omega}=\nabla\times\vec{u}$,
since Eqn.~\ref{eqn:circ-def} can be rewritten,
using Stokes' theorem, as
\begin{linenomath*}
\begin{equation}
    \Gamma=\int_A \vec{\omega}\cdot d\vec{A},
\end{equation}
\end{linenomath*}
which is an area integral over any surface $A$ 
whose boundary is the closed curve $C$, 
and where
$d\vec{A}$ is a surface area element whose vector direction
is perpendicular to the surface. 
Helmholtz \cite{helmholtz1858integrals}
had earlier formulated some related conservation statements
in terms of vorticity.
An associated quantity is
the potential vorticity (PV), which is the dot product of the
vorticity vector and the 
entropy gradient, $\nabla s$,
weighted by the fluid density $\rho$;
that is,
\begin{linenomath*}
\begin{equation}
    \label{eq:pv_def}
    PV=\frac{\vec{\omega}\cdot\nabla s}{\rho}.
\end{equation}
\end{linenomath*}
Ertel \cite{ertel1942neuer}
showed that PV is conserved following the trajectory
of a fluid parcel, which builds on an earlier result of
Rossby \cite{rossby1940planetary}
for the shallow water equations.

PV is a fundamental quantity of interest in geophysical flows
\cite{muller1995ertel,salmon1998lectures}.
It is extensively used in analysis of weather events \cite{davis1991potential,lackmann2002cold}, the general ocean circulation \cite{rhines1986vorticity, pollard1990large, holland1984dynamics, muller1995ertel}, magnetohydrodynamics (MHD) and astrophysical fluid dynamics \cite{webb2015potential}. The wide-ranging applications of PV stem from its conservation properties and the notion of PV inversion \cite{hoskins1985use,smith2017precipitating,wetzel2020potential}. 

Despite the importance of PV and the further knowledge gained over approximately a century, the conservation laws of circulation and PV are limited to fluids that are relatively idealized. In particular, these conservation laws apply to
an ocean or atmosphere that is composed of a single fluid substance---water in the ocean, or dry air in the atmosphere. In nature, the ocean and atmosphere include other substances which are of fundamental significance. Moisture and clouds in the atmosphere are associated with weather events such as hurricanes, tornadoes, and monsoons. Salinity in the ocean affects the density and helps drive the thermohaline circulation and meridional overturning circulation that move heat from low latitudes to high latitudes \cite{marshall1999open,curry2005dilution,peterson2006trajectory,vallis2017atmospheric}. As one route to better understanding and predicting these phenomena, it is desirable to have knowledge of the associated conserved quantities.

The main goal of the present paper is to 
derive conservation principles for circulation and
potential vorticity, for the case of an ocean with salinity
or an atmosphere with moisture and clouds.
The atmospheric case will be described first,
and then the oceanic case will follow in a similar way.
Also, circulation will be treated first, and 
then potential vorticity.

\section{Circulation}

The starting point is the evolution equation for the
circulation $\Gamma$, which was defined above in
Eqn.~\ref{eqn:circ-def}.
Its evolution equation follows from the equations for
conservation of mass and momentum
\cite{thorpe2003bjerknes,bjerknes1898hydrodynamischen,vallis2017atmospheric}
and it takes the form
\begin{linenomath*}
\begin{equation}
\label{eq:circulation_evol}
    \frac{D\Gamma}{Dt}=\oint_{C(t)} \frac{1}{\rho} \nabla p\cdot d\vec{l},
\end{equation}
\end{linenomath*}
where $p$ is the pressure, 
$C(t)$ is a closed curve that moves along with the fluid,
and $D/Dt=\partial/\partial t + \vec{u}\cdot\nabla$ 
is a material derivative---i.e., a derivative
along the trajectory of a fluid parcel. 
See the Supplementary Materials for a derivation. 
The fluid is 
assumed to be compressible and inviscid.  

Before considering the moist case, it is instructive
to consider the dry case, for which it is known that
circulation is conserved, under certain assumptions
on the curve $C(t)$. To see this, one can rewrite (\ref{eq:circulation_evol}) as
\begin{linenomath*}
\begin{equation}
\label{eq:theta_dpi}
    \frac{D\Gamma}{Dt} = \oint_{C(t)} \theta\; d\pi,
\end{equation}
\end{linenomath*}
where the Exner function $\pi(p)$ is a function of pressure,
and where the potential temperature $\theta$ is a function
of entropy, as $s=c_p\log \theta +const.$, where
$c_p$, commonly assumed constant for a dry atmosphere,
is the specific heat at constant pressure.
The details of this derivation are provided in the Supplementary Materials. 
From (\ref{eq:theta_dpi}), one can deduce the scenarios
when the right-hand side is zero and circulation $\Gamma$
is conserved. In particular, if $\theta$ (and hence also
entropy $s$) is a constant on the closed curve $C(t)$,
then the right-hand side of (\ref{eq:theta_dpi}) is zero,
by the fundamental theorem of calculus for line integrals.
Consequently, circulation is conserved,
\begin{linenomath*}
\begin{equation}
\label{eq:dry_kelvin}
    \frac{D\Gamma}{Dt} = 0,
\end{equation}
\end{linenomath*}
if the closed curve
$C(t)$ lies on a surface of constant entropy. See Figure~\ref{fig:dry-circulation}
for an illustration.
In order for the material curve $C(t)$ to remain on a surface of constant entropy
for all times, the fluid is assumed to be adiabatic, so that $Ds/Dt=0$. 
Finally, from (\ref{eq:dry_kelvin}), the circulation $\Gamma$ is then said to be
a \emph{material invariant}, since its material derivative is zero.

In the moist case, one could try to repeat the derivation from the dry case to arrive at a circulation theorem. However, this line of reasoning fails for a moist system, since the integrand that would arise in (\ref{eq:theta_dpi}) would no longer be a material invariant, due to phase changes and cloud latent heating. 

\begin{figure}
    \centering
    \includegraphics[width=0.8\textwidth]{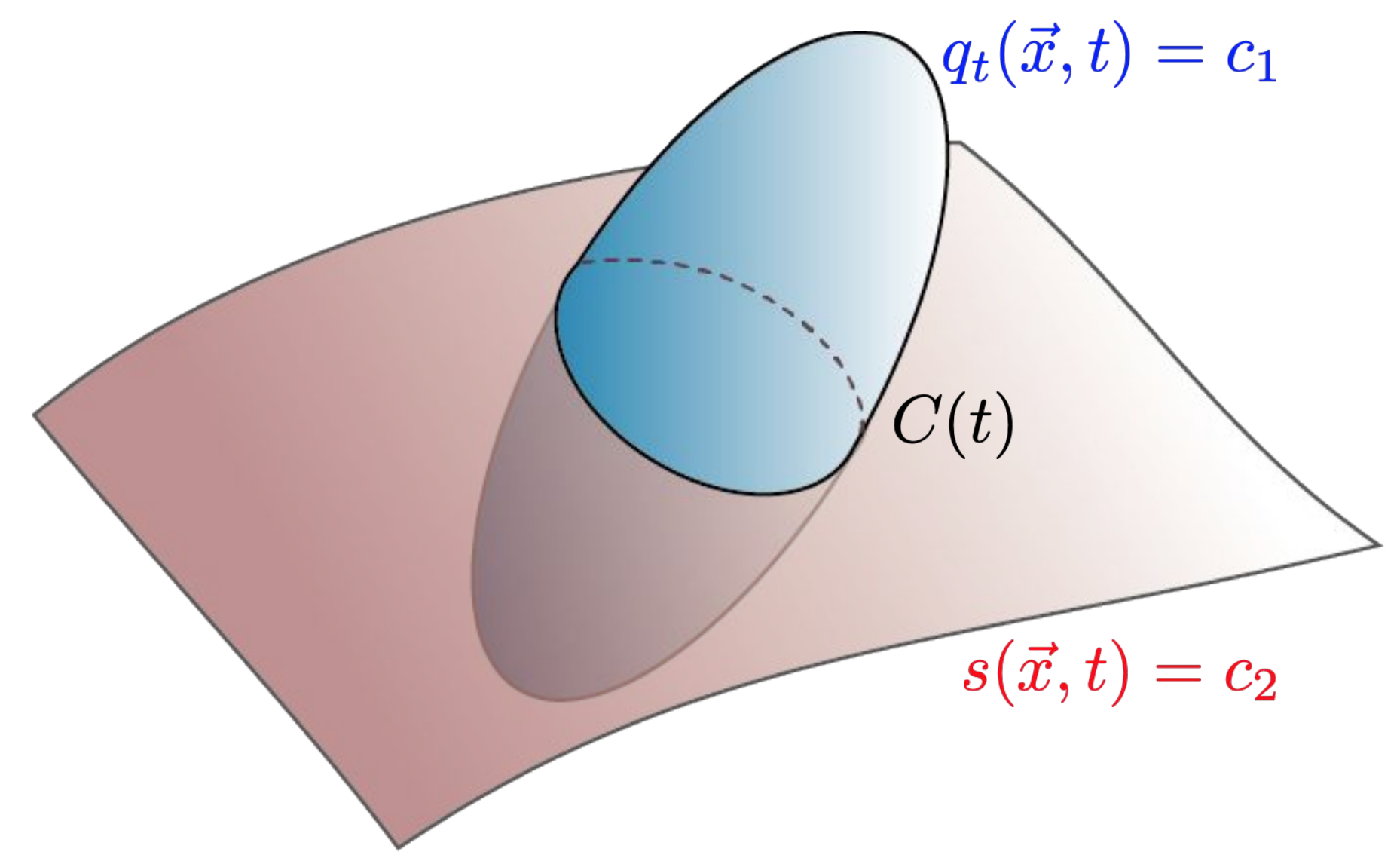}
    \caption{Geometry of the circulation theorem for an atmosphere with clouds
    or an ocean with salinity. The curve $C(t)$ is a curve of constant entropy $s$ and 
    total water specific humidity $q_t$.}
    \label{fig:moist-circulation}
\end{figure}

We now show that
the challenges of the moist case can actually be overcome
by referring to fundamental principles of
moist thermodynamics. Two derivations will be shown:
a derivation that directly refers to enthalpy,
and a derivation that does not.
In terms of the enthalpy $h(s,p,q_t)$, 
the fundamental thermodynamic relation
\cite{landau1980statistical,landau1987fluid,p08} 
states that
\begin{linenomath*}
\begin{equation}
dh=T\,ds+\rho^{-1}\ dp + \mu\,dq_t,
\label{eqn:first-law-standard}
\end{equation}
\end{linenomath*}
which can be viewed as the differential of the function
$h(s,p,q_t)$ as
\begin{linenomath*}
\begin{equation}
dh
=\left(\frac{\partial h}{\partial s}\right)_{p,q_t}\,ds
+\left(\frac{\partial h}{\partial p}\right)_{s,q_t}\ dp 
+\left(\frac{\partial h}{\partial q_t}\right)_{s,p}\,dq_t,
\label{eqn:first-law-partials}
\end{equation}
\end{linenomath*}
where $h$ is enthalpy, $T$ is temperature,
$q_t$ is the total water specific humidity, 
and $\mu=\mu_v-\mu_d$ is the difference of the 
chemical potentials 
associated with water vapor and dry air, respectively.
As a first important observation, notice that 
(\ref{eqn:first-law-standard})
includes the term $\rho^{-1}\; dp$,
which is also the sole term
that appears in the evolution equation
for circulation in (\ref{eq:circulation_evol}).
Therefore, inserting the fundamental relation (\ref{eqn:first-law-standard}) 
into the evolution equation for circulation in (\ref{eq:circulation_evol}), we have
\begin{linenomath*}
\begin{equation}
\label{eqn:circ-s-qt}
    \frac{D\Gamma}{Dt} = - \oint_{C(t)} T\; ds - \oint_{C(t)} \mu\; dq_t,
\end{equation}
\end{linenomath*}
where it was used that $\oint_{C(t)} dh=0$ for a closed curve.
The aim now is to identify scenarios when the right-hand side is zero.
In this direction, note that the line integrals are zero if the curve $C(t)$ lies
on a surface of constant entropy $s$ \emph{and} a surface of constant $q_t$.
Furthermore, in order for the material curve to have constant $s$ and $q_t$
for all times, it is assumed that $Ds/Dt=0$ and $Dq_t/Dt=0$,
so that $s$ and $q_t$ are both material invariants, if reversible phase changes
are assumed between water vapor and liquid water, and precipitation is absent.
Therefore, the right hand side of (\ref{eqn:circ-s-qt}) vanishes and we have
\begin{linenomath*}
\begin{equation}
\label{eq:moist_kelvin}
\frac{D\Gamma}{Dt} = 0,
\end{equation}
\end{linenomath*}
if $C(t)$ is a closed curve of constant $s$ and $q_t$.  
See Figure~\ref{fig:moist-circulation} for an illustration.
This is the moist, cloudy analog of the circulation theorems of Kelvin and Bjerknes.

As an alternative derivation of (\ref{eq:moist_kelvin}) that does not 
directly refer to enthalpy, return again to the starting point of 
(\ref{eq:circulation_evol}), where the integrand is $\rho^{-1}$.
Now use a fundamental property of moist thermodynamics:
any moist thermodynamic variable can be written as a function of $s$, $q_t$ and $p$ \cite{p08}. 
Hence, $\rho^{-1}\; dp$ can be written as $\rho^{-1}(s,q_t,p)\; dp$, which, 
on a circuit $C(t)$ of constant entropy $s$ and total water mixing ratio $q_t$, 
is a function of pressure $p$ alone. Consequently, (\ref{eq:circulation_evol})
can furthermore be simplified to
\begin{linenomath*}
\begin{equation}
    \label{eq:circ-theorem-df}
    \frac{D \Gamma}{Dt} = \oint_{C(t)} df,
\end{equation}
\end{linenomath*}
where $f(p)$ is an antiderivative that satisfies 
$df=(df/dp)\; dp=\rho^{-1}(s,q_t,p)\; dp$.
By (\ref{eqn:first-law-standard}) and (\ref{eqn:first-law-partials}),
since $(dh/dp)_{s,q_t}=\rho^{-1}$,
the function $f$ can be identified as the enthalpy $h$,
which is insightful but not essential for this derivation.
Finally, since $C(t)$ is a closed curve, it follows from (\ref{eq:circ-theorem-df}) that
\begin{linenomath*}
\begin{equation}
    \label{eq:circ-theorem}
    \frac{D \Gamma}{Dt} = 0,
\end{equation}
\end{linenomath*}
if $C(t)$ is a closed curve of constant $s$ and $q_t$. 

As a comparison of the dry and moist cases, notice that both apply to 
material curves $C(t)$ on surfaces of constant entropy $s$.
However, while \emph{any} such curve is valid for a dry atmosphere,
the curve must also have constant $q_t$ for a moist, cloudy atmosphere.

\section{Potential vorticity}

Now consider the second quantity of interest here,
defined above in (\ref{eq:pv_def}):
potential vorticity. 
Its evolution equation follows from the equations for
conservation of mass, momentum, and entropy \cite{muller1995ertel,schubert2001potential}
and it takes the form 
\begin{linenomath*}
\begin{equation}
    \rho \frac{D}{Dt} \left( \frac{\vec{\omega} \cdot \nabla s} {\rho} \right) =  \nabla s \cdot \nabla \times \left( \frac{1}{\rho} \nabla p\right)
    \label{eqn:pv-evol}
\end{equation}
\end{linenomath*}
See the Supplementary Materials for a derivation.
On the right-hand side, the term $\nabla\times(\rho^{-1}\nabla p)$
arises from the curl of the pressure gradient in the 
statement of conservation of momentum.

For a dry atmosphere, PV is conserved for each fluid parcel,
and it is helpful to review this case as a prelude
to the moist case.
To see how dry PV is conserved, note that the right-hand side of (\ref{eqn:pv-evol})
can be written as
\begin{linenomath*}
\begin{equation}
    \nabla s \cdot \nabla \times \left( \frac{1}{\rho} \nabla p\right)
    =
    -\frac{1}{\rho^2}\nabla s \cdot \nabla p \times \nabla\rho ,
    \label{eqn:triple-prod}
\end{equation}
\end{linenomath*}
which is a scalar triple product. Then one can use a fundamental
property of thermodynamics for a dry atmosphere:
the entropy can be expressed as a function of pressure
and density, so that
$s=s(p,\rho)$.
Consequently, we have
\begin{linenomath*}
\begin{equation}
    \nabla s 
    = \frac{\partial s}{\partial p}\nabla p 
    +\frac{\partial s}{\partial \rho}\nabla \rho
\end{equation}
\end{linenomath*}
and inserting this into (\ref{eqn:triple-prod})
shows that the right-hand side of (\ref{eqn:pv-evol}) is zero,
and the PV is conserved for each fluid parcel:
\begin{linenomath*}
\begin{equation}
    \frac{D}{Dt} \left( \frac{\vec{\omega} \cdot \nabla s} {\rho} \right) = 0.
\end{equation}
\end{linenomath*}
In essence, this proof works smoothly because the entropy $s$
is a function of pressure $p$ and density $\rho$,
for a dry atmosphere.
However, for a moist atmosphere, due to the effects of
phase changes and cloud latent heating, it is not true that
$s=s(p,\rho)$, and it is therefore difficult to see how to
extend these proof ideas to apply to the moist case.

For a moist atmosphere with clouds, we now show that
the difficulty of PV conservation can be resolved
by appealing to fundamentals of moist thermodynamics.
In particular, use the gradient form of the fundamental
thermodynamic relation in (\ref{eqn:first-law-standard}),
and insert into (\ref{eqn:pv-evol}) for the $\rho^{-1}\nabla p$ term.
Then, after using vector calculus identities, (\ref{eqn:pv-evol}) becomes
\begin{linenomath*}
\begin{equation}
    \rho \frac{D}{Dt} \left( \frac{\vec{\omega} \cdot \nabla s} {\rho} \right) =  -\nabla s \cdot \nabla \times \left( \mu \nabla q_t\right).
    \label{eqn:pv-evol-mu-qt}
\end{equation}
\end{linenomath*}
The benefit of having replaced $\rho^{-1}\nabla p$ with 
$\mu\nabla q_t$ will be seen below
and arises from the
fact that the right-hand side of (\ref{eqn:pv-evol-mu-qt})
now involves two quantities, $s$ and $q_t$,
which are both conserved for each fluid parcel,
for the case of reversible phase changes between
water vapor and cloud liquid water, with no precipitation.

To proceed with finding the conservation law for moist PV, 
we abandon the idea that moist PV
can be conserved for each fluid parcel. 
In other words, we abandon hope of showing that the
right-hand side of (\ref{eqn:pv-evol-mu-qt}) is zero.
Instead, consider the integral of
moist PV over certain volumes. The evolution equation
of volume-integrated PV is
\begin{linenomath*}
\begin{equation}
\label{eqn:pv-line-1}
\frac{D}{Dt} \int \int \int_{V(t)} \frac{\vec{\omega} \cdot \nabla s}{\rho} \; \rho\, dV
=  \int \int \int_{V(t)} \rho \frac{D}{Dt} \left(\frac{\vec{\omega} \cdot \nabla s}{\rho} \right) \; dV
\end{equation}
\end{linenomath*}
\begin{linenomath*}
\begin{equation}
\label{eqn:pv-line-2}
     = -\int \int \int_{V(t)} \nabla \cdot \left(\mu\nabla q_t \times \nabla s\right) \; dV
\end{equation}
\end{linenomath*}
\begin{linenomath*}
\begin{equation}
\label{eqn:pv-line-3}
     = -\int \int_{A(t)} \mu\nabla q_t \times \nabla s \cdot d\vec{A},
     \label{eqn:int-pv-evol-surf}
\end{equation}
\end{linenomath*}
where the volume integral in (\ref{eqn:pv-line-1}) is mass-weighted using density $\rho$,
and where the volume $V(t)$ is a material volume that moves with the fluid
but otherwise remains to be specified. 
See the Supplementary Materials for details of the derivation.
In brief, (\ref{eqn:pv-line-1}) uses the transport theorem, 
(\ref{eqn:pv-line-2}) uses (\ref{eqn:pv-evol-mu-qt}) and some vector calculus
identities, and (\ref{eqn:pv-line-3}) follows from the divergence theorem,
where $A(t)$ is the surface that encloses the material volume $V(t)$.

Now it is time to identify the special volumes
for which (\ref{eqn:int-pv-evol-surf}) is zero
and integrated moist PV is conserved.
The intuition is similar to the case of the moist
circulation theorem in (\ref{eqn:circ-s-qt})
where the appearance of the line elements $ds$ and $dq_t$
suggested the choice of curves of constant $s$ and $q_t$.
Here, it is a surface integral that arises in
(\ref{eqn:int-pv-evol-surf}), and in order for the
integral to vanish, the surface area element
$d\vec{A}$ must be perpendicular to $\nabla s$ or $\nabla q_t$.
Hence, we have
\begin{linenomath*}
\begin{equation}
\label{eqn:moist-pv}
\frac{D}{Dt} \int \int \int_{V(t)} \frac{\vec{\omega} \cdot \nabla s}{\rho} \; \rho\, dV
=  0,
\end{equation}
\end{linenomath*}
if the volume is defined to be enclosed by surfaces of
constant $s$ and surfaces of constant $q_t$.
An alternative derivation, which does not explicitly use the 
fundamental thermodynamic relation,
is shown in the Supplementary Materials.
Either derivation leads to the same result, (\ref{eqn:moist-pv}),
the conservation law for PV for a cloudy atmosphere,
which applies to volume-integrated PV.

\begin{figure}
    \centering
    \includegraphics[width=\textwidth]{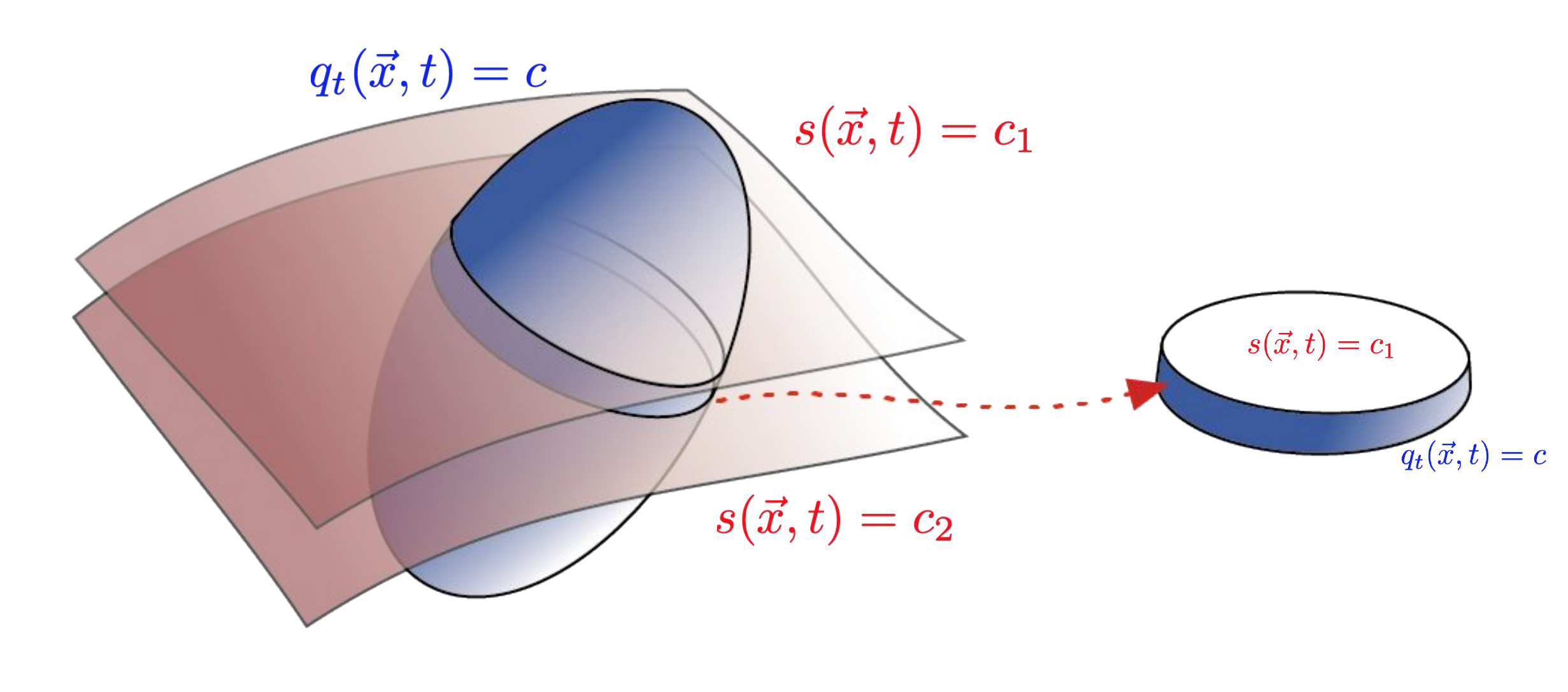}
    \caption{Geometry of the PV conservation law for an atmosphere with clouds 
    or an ocean with salinity.
    The conserved quantity is not PV of individual parcels, but PV integrated over 
    certain cylinder-like volumes.}
    \label{fig:moist-pv}
\end{figure}

More generally, the surface could be composed piecewise, based on level surfaces
of general functions $F(s,q_t)$, since any such level surface is perpendicular to
$\nabla s\times\nabla q_t$. Nevertheless, it is convenient and simple to choose
surfaces of constant $s$ or constant $q_t$, as illustrated in Figure~\ref{fig:moist-pv}.
The simplest such volume is shaped like a thin cylinder, or pill box, or pancake.

As perhaps the most special case, notice that the volume $V(t)$ 
in Figure~\ref{fig:moist-pv} will shrink to a point
at a maximum or minimum value of $q_t$ on a constant-entropy surface.
As a result, at minima or maxima of $q_t$, the PV is conserved for the
individual parcel.

As another generalization, notice that the PV could be defined as
$\rho^{-1}\vec{\omega}\cdot\nabla q_t$ with $q_t$ in the role that is normally
played by entropy $s$. A similar conservation law also follows for such a
$q_t$-based PV. Moreover, the PV could be defined as 
$\rho^{-1}\vec{\omega}\cdot\nabla F(s,q_t)$ for any function $F$ of $s$ and $q_t$.
In that case, the proof proceeds in the same way by writing the 
fundamental thermodynamic relation
in a form that replaces $s$ and $q_t$ by $F(s,q_t)$ and $G(s,q_t)$,
where $G$ is arbitrary but functionally independent of $F$.

For an ocean with salinity, the conservation laws for circulation and PV 
follow in the same way. Simply replace specific humidity $q_t$ 
by salinity $S$, and to avoid confusion, change the 
notation of the entropy $s$ to instead be $\eta$, 
as is common in oceanography.
Table~\ref{tab:summary} summarizes the conservation laws 
for both the cloudy atmosphere and salty ocean.

\begin{table}
\begin{tabular}{lll}
\hline\hline
\\[-6pt]
& Cloudy Atmosphere & Salty Ocean
\\[4pt]
\hline
\\
Circulation Theorem & $\frac{D}{Dt}\oint_{C(t)} \vec{u}\cdot d\vec{l}=0$ & $\frac{D}{Dt}\oint_{C(t)} \vec{u}\cdot d\vec{l}=0$ 
\\[11pt]
Curve $C(t)$ & Constant $s$ and $q_t$ & Constant $\eta$ and $S$ 
\\[11pt]
PV Conservation Law 
& $\frac{D}{Dt} \int \int \int_{V(t)} \frac{\vec{\omega} \cdot \nabla s}{\rho} \; \rho\, dV =  0$ 
& $\frac{D}{Dt} \int \int \int_{V(t)} \frac{\vec{\omega} \cdot \nabla \eta}{\rho} \; \rho\, dV =  0$
\\[11pt]
Surface of Volume $V(t)$ & Constant $s$ or $q_t$ & Constant $\eta$ or $S$
\\[11pt]
\hline
\end{tabular}
\caption{\textbf{Summary of conservation laws} Conservation laws for an atmosphere with clouds and an ocean
with salinity. Commonly used notation for a cloudy atmosphere is
$s$ and $q_t$ for the
entropy and total water specific humidity, respectively,
and for a salty ocean is $\eta$ and $S$ for the entropy and salinity, respectively.
Rotation and the Coriolis force could also be incorporated
with modifications described in (\ref{eqn:circ-thm-rot}) and 
(\ref{eqn:abs-vort}).
}
\label{tab:summary}
\end{table}

\section{Symmetry--conservation relationship}
A symmetry--conservation relationship is also associated with these conservation laws
of circulation and PV. In particular, it is a particle-relabeling symmetry, which,
as the name suggests, involves particles whose positions are 
$\vec{x}(\vec{a},t)$ or the relabeled positions $\vec{x}(\vec{a}\,^\prime,t)$,
where the original label $\vec{a}$ was replaced by the relabeling $\vec{a}\,^\prime=\vec{a}\,^\prime (\vec{a},t)$.
In a discrete setting, the relabeling of $\vec{x}_n(t)$ to $\vec{x}_{n'}(t)$
would simply involve a permutation of the labels, $n=1,2,3,\cdots, N$,
but there is no guarantee of an associated conservation law for such a discrete symmetry.
For the continuum setting of a cloudy atmosphere, the symmetry applies to the 
Lagrangian
\begin{linenomath*}
\begin{equation}
\mathcal{L}(t) = \int\int\int \rho \left[ \frac{1}{2}|\vec{u}|^2 - E(\rho,s,q_t) - \Phi(\vec{x}) \right] \; dV_a,
\label{eqn:lag}
\end{equation}
\end{linenomath*}
where $E(\rho,s,q_t)$ is the internal energy, $\Phi(\vec{x})$ is the gravitational potential,
and the integral is over the particle labels $\vec{a}$, as the continuum version of
a sum over the particles. 

A key observation is that only some special particle relabelings will leave the Lagrangian unchanged.
For instance, for a dry atmosphere, the internal energy $E(\rho,s)$ is a function
of density and entropy, and it is well-known that 
$E(\rho,s)$ and the Lagrangian remain unchanged
for certain particle relabelings on surfaces of constant entropy $s$ 
\cite{salmon1998lectures};
and it follows that the Kelvin--Bjerknes circulation theorem holds for
curves $C(t)$ that lie on surfaces of constant entropy, as mentioned in
(\ref{eq:dry_kelvin}) and Figure~\ref{fig:dry-circulation} above.
Now, for a moist atmosphere, one can see the natural extension:
the particle relabeling should keep both $s$ and $q_t$ unchanged in order to keep
$E(\rho,s,q_t)$ and the Lagrangian unchanged; consequently, the moist circulation theorem
in (\ref{eq:moist_kelvin}) and Figure~\ref{fig:moist-circulation} holds for curves $C(t)$
of constant $s$ and $q_t$. 
These symmetry--conservation ideas have been derived in detail for a moist Boussinesq
setting 
\cite{kooloth2022hamiltons}
and here provide motivation and further understanding of the moist conservation laws
of circulation and PV.
For instance, 
because the particle relabeling is constrained
to curves $C(t)$ of both constant $s$ and $q_t$,
it follows that moist PV is not conserved for individual parcels.

\section{Concluding discussion}

Rotation and the Coriolis term could also be incorporated here, 
in which case the circulation theorem becomes
\begin{linenomath*}
\begin{equation}
    \frac{D}{Dt}\oint_{C(t)} (\vec{u}+\vec{\Omega}\times\vec{x})\cdot d\vec{l}=0,
    \label{eqn:circ-thm-rot}
\end{equation}
\end{linenomath*}
where $\Omega$ is the angular rotation vector
and $\vec{x}$ is the position vector.
The extension to include rotation follows standard procedures
as in cases without clouds or salinity
\cite{salmon1998lectures,vallis2017atmospheric,cotter2014variational}.
Furthermore, the PV conservation laws have the same form 
as in Table~\ref{tab:summary} except with
$\omega=\nabla\times\vec{u}$ replaced by absolute vorticity,
\begin{linenomath*}
\begin{equation}
    \vec{\omega}_a = \nabla\times\vec{u} + 2\vec{\Omega},
    \label{eqn:abs-vort}
\end{equation}
\end{linenomath*}
which is the sum of the relative vorticity and the
angular rotation vector.


Many applications are possible for the conservation laws presented here.
One common use of PV conservation is to diagnose non-conservation,
which is an indication of additional physical processes
\cite{haynes1987evolution}.
The PV conservation laws here could be used for similar
purposes, with certain modifications. For instance, 
cloud latent heating is considered one of the sources of
non-conservation for dry PV, but is incorporated into the
conservation law here for moist PV.
Another application of PV conservation is in numerical models of weather and climate.
It is often desirable to design such numerical methods to accurately satisfy the
PV conservation law 
\cite{thuburn2008some,taylor2010compatible}.
The conservation laws here provide additional targets for numerical conservation,
now incorporating additional physical processes of salinity and cloud latent heating.

\section{Open Research}
Data sharing not applicable to this article as no data sets were generated or analysed during the current study.






\acknowledgments
Leslie Smith and Sam Stechmann gratefully acknowledge grant support from the Division of Mathematical Sciences of the National Science Foundation, in conjunction with the award DMS-1907667.


\bibliography{references}

\begin{thebibliography}{}

\bibitem [\protect \citeauthoryear {%
Bjerknes%
}{%
Bjerknes%
}{%
{\protect \APACyear {1898}}%
}]{%
bjerknes1898hydrodynamischen}
\APACinsertmetastar {%
bjerknes1898hydrodynamischen}%
\begin{APACrefauthors}%
Bjerknes, V.%
\end{APACrefauthors}%
\unskip\
\newblock
\APACrefYearMonthDay{1898}{}{}.
\newblock
{\BBOQ}\APACrefatitle {{\"U}ber einen hydrodynamischen Zirkulationssatz und
  seine Anwendung auf die Mechanik der Atmosph{\"a}re und des Weltmeeres}
  {{\"U}ber einen hydrodynamischen zirkulationssatz und seine anwendung auf die
  mechanik der atmosph{\"a}re und des weltmeeres}.{\BBCQ}
\newblock
\APACjournalVolNumPages{Kongl. Svenska Vetenskapsakad. Handl.
  Bd}{31}{}{97--106}.
\PrintBackRefs{\CurrentBib}

\bibitem [\protect \citeauthoryear {%
Cotter%
\ \BBA {} Holm%
}{%
Cotter%
\ \BBA {} Holm%
}{%
{\protect \APACyear {2014}}%
}]{%
cotter2014variational}
\APACinsertmetastar {%
cotter2014variational}%
\begin{APACrefauthors}%
Cotter, C.%
\BCBT {}\ \BBA {} Holm, D.%
\end{APACrefauthors}%
\unskip\
\newblock
\APACrefYearMonthDay{2014}{}{}.
\newblock
{\BBOQ}\APACrefatitle {Variational formulations of sound-proof models}
  {Variational formulations of sound-proof models}.{\BBCQ}
\newblock
\APACjournalVolNumPages{Quart. J. Roy. Met. Soc.}{140}{683}{1966--1973}.
\PrintBackRefs{\CurrentBib}

\bibitem [\protect \citeauthoryear {%
Curry%
\ \BBA {} Mauritzen%
}{%
Curry%
\ \BBA {} Mauritzen%
}{%
{\protect \APACyear {2005}}%
}]{%
curry2005dilution}
\APACinsertmetastar {%
curry2005dilution}%
\begin{APACrefauthors}%
Curry, R.%
\BCBT {}\ \BBA {} Mauritzen, C.%
\end{APACrefauthors}%
\unskip\
\newblock
\APACrefYearMonthDay{2005}{}{}.
\newblock
{\BBOQ}\APACrefatitle {Dilution of the northern {North Atlantic Ocean} in
  recent decades} {Dilution of the northern {North Atlantic Ocean} in recent
  decades}.{\BBCQ}
\newblock
\APACjournalVolNumPages{Science}{308}{5729}{1772--1774}.
\PrintBackRefs{\CurrentBib}

\bibitem [\protect \citeauthoryear {%
Davis%
\ \BBA {} Emanuel%
}{%
Davis%
\ \BBA {} Emanuel%
}{%
{\protect \APACyear {1991}}%
}]{%
davis1991potential}
\APACinsertmetastar {%
davis1991potential}%
\begin{APACrefauthors}%
Davis, C\BPBI A.%
\BCBT {}\ \BBA {} Emanuel, K\BPBI A.%
\end{APACrefauthors}%
\unskip\
\newblock
\APACrefYearMonthDay{1991}{}{}.
\newblock
{\BBOQ}\APACrefatitle {Potential vorticity diagnostics of cyclogenesis}
  {Potential vorticity diagnostics of cyclogenesis}.{\BBCQ}
\newblock
\APACjournalVolNumPages{Monthly Weather Review}{119}{8}{1929--1953}.
\PrintBackRefs{\CurrentBib}

\bibitem [\protect \citeauthoryear {%
Ertel%
}{%
Ertel%
}{%
{\protect \APACyear {1942}}%
}]{%
ertel1942neuer}
\APACinsertmetastar {%
ertel1942neuer}%
\begin{APACrefauthors}%
Ertel, H.%
\end{APACrefauthors}%
\unskip\
\newblock
\APACrefYearMonthDay{1942}{}{}.
\newblock
{\BBOQ}\APACrefatitle {Ein neuer hydrodynamischer Wirbelsatz} {Ein neuer
  hydrodynamischer wirbelsatz}.{\BBCQ}
\newblock
\APACjournalVolNumPages{Met. Zeitschr.}{59}{}{277--281}.
\PrintBackRefs{\CurrentBib}

\bibitem [\protect \citeauthoryear {%
Haynes%
\ \BBA {} McIntyre%
}{%
Haynes%
\ \BBA {} McIntyre%
}{%
{\protect \APACyear {1987}}%
}]{%
haynes1987evolution}
\APACinsertmetastar {%
haynes1987evolution}%
\begin{APACrefauthors}%
Haynes, P\BPBI H.%
\BCBT {}\ \BBA {} McIntyre, M\BPBI E.%
\end{APACrefauthors}%
\unskip\
\newblock
\APACrefYearMonthDay{1987}{}{}.
\newblock
{\BBOQ}\APACrefatitle {On the evolution of vorticity and potential vorticity in
  the presence of diabatic heating and frictional or other forces} {On the
  evolution of vorticity and potential vorticity in the presence of diabatic
  heating and frictional or other forces}.{\BBCQ}
\newblock
\APACjournalVolNumPages{J. Atmos. Sci.}{44}{5}{828--841}.
\PrintBackRefs{\CurrentBib}

\bibitem [\protect \citeauthoryear {%
Helmholtz%
}{%
Helmholtz%
}{%
{\protect \APACyear {1858}}%
}]{%
helmholtz1858integrals}
\APACinsertmetastar {%
helmholtz1858integrals}%
\begin{APACrefauthors}%
Helmholtz, H\BPBI v.%
\end{APACrefauthors}%
\unskip\
\newblock
\APACrefYearMonthDay{1858}{}{}.
\newblock
{\BBOQ}\APACrefatitle {{\"U}ber Integrale der hydrodynamischen Gleichungen,
  welche den Wirbelbewegungen entsprechen} {{\"U}ber integrale der
  hydrodynamischen gleichungen, welche den wirbelbewegungen
  entsprechen}.{\BBCQ}
\newblock
\APACjournalVolNumPages{J. Angew. Math.}{55}{}{25-55}.
\PrintBackRefs{\CurrentBib}

\bibitem [\protect \citeauthoryear {%
Holland%
, Keffer%
\BCBL {}\ \BBA {} Rhines%
}{%
Holland%
\ \protect \BOthers {.}}{%
{\protect \APACyear {1984}}%
}]{%
holland1984dynamics}
\APACinsertmetastar {%
holland1984dynamics}%
\begin{APACrefauthors}%
Holland, W.%
, Keffer, T.%
\BCBL {}\ \BBA {} Rhines, P.%
\end{APACrefauthors}%
\unskip\
\newblock
\APACrefYearMonthDay{1984}{}{}.
\newblock
{\BBOQ}\APACrefatitle {Dynamics of the oceanic general circulation: the
  potential vorticity field} {Dynamics of the oceanic general circulation: the
  potential vorticity field}.{\BBCQ}
\newblock
\APACjournalVolNumPages{Nature}{308}{5961}{698--705}.
\PrintBackRefs{\CurrentBib}

\bibitem [\protect \citeauthoryear {%
Hoskins%
, McIntyre%
\BCBL {}\ \BBA {} Robertson%
}{%
Hoskins%
\ \protect \BOthers {.}}{%
{\protect \APACyear {1985}}%
}]{%
hoskins1985use}
\APACinsertmetastar {%
hoskins1985use}%
\begin{APACrefauthors}%
Hoskins, B\BPBI J.%
, McIntyre, M\BPBI E.%
\BCBL {}\ \BBA {} Robertson, A\BPBI W.%
\end{APACrefauthors}%
\unskip\
\newblock
\APACrefYearMonthDay{1985}{}{}.
\newblock
{\BBOQ}\APACrefatitle {On the use and significance of isentropic potential
  vorticity maps} {On the use and significance of isentropic potential
  vorticity maps}.{\BBCQ}
\newblock
\APACjournalVolNumPages{Quart. J. Roy. Met. Soc.}{111}{470}{877--946}.
\PrintBackRefs{\CurrentBib}

\bibitem [\protect \citeauthoryear {%
Kooloth%
, Smith%
\BCBL {}\ \BBA {} Stechmann%
}{%
Kooloth%
\ \protect \BOthers {.}}{%
{\protect \APACyear {2022}}%
}]{%
kooloth2022hamiltons}
\APACinsertmetastar {%
kooloth2022hamiltons}%
\begin{APACrefauthors}%
Kooloth, P.%
, Smith, L\BPBI M.%
\BCBL {}\ \BBA {} Stechmann, S\BPBI N.%
\end{APACrefauthors}%
\unskip\
\newblock
\APACrefYearMonthDay{2022}{}{}.
\newblock
{\BBOQ}\APACrefatitle {Hamilton's principle with phase changes and conservation
  principles for moist potential vorticity} {Hamilton's principle with phase
  changes and conservation principles for moist potential vorticity}.{\BBCQ}
\newblock
\APACjournalVolNumPages{}{}{}{submitted}.
\PrintBackRefs{\CurrentBib}

\bibitem [\protect \citeauthoryear {%
Lackmann%
}{%
Lackmann%
}{%
{\protect \APACyear {2002}}%
}]{%
lackmann2002cold}
\APACinsertmetastar {%
lackmann2002cold}%
\begin{APACrefauthors}%
Lackmann, G\BPBI M.%
\end{APACrefauthors}%
\unskip\
\newblock
\APACrefYearMonthDay{2002}{}{}.
\newblock
{\BBOQ}\APACrefatitle {Cold-frontal potential vorticity maxima, the low-level
  jet, and moisture transport in extratropical cyclones} {Cold-frontal
  potential vorticity maxima, the low-level jet, and moisture transport in
  extratropical cyclones}.{\BBCQ}
\newblock
\APACjournalVolNumPages{Monthly Weather Review}{130}{1}{59--74}.
\PrintBackRefs{\CurrentBib}

\bibitem [\protect \citeauthoryear {%
Landau%
\ \BBA {} Lifshitz%
}{%
Landau%
\ \BBA {} Lifshitz%
}{%
{\protect \APACyear {1980}}%
}]{%
landau1980statistical}
\APACinsertmetastar {%
landau1980statistical}%
\begin{APACrefauthors}%
Landau, L\BPBI D.%
\BCBT {}\ \BBA {} Lifshitz, E\BPBI M.%
\end{APACrefauthors}%
\unskip\
\newblock
\APACrefYear{1980}.
\newblock
\APACrefbtitle {Statistical Physics, Part 1} {Statistical physics, part 1}\
  (\PrintOrdinal{3}\ \BEd, \BVOL~5).
\newblock
\APACaddressPublisher{}{Elsevier}.
\PrintBackRefs{\CurrentBib}

\bibitem [\protect \citeauthoryear {%
Landau%
\ \BBA {} Lifshitz%
}{%
Landau%
\ \BBA {} Lifshitz%
}{%
{\protect \APACyear {1987}}%
}]{%
landau1987fluid}
\APACinsertmetastar {%
landau1987fluid}%
\begin{APACrefauthors}%
Landau, L\BPBI D.%
\BCBT {}\ \BBA {} Lifshitz, E\BPBI M.%
\end{APACrefauthors}%
\unskip\
\newblock
\APACrefYear{1987}.
\newblock
\APACrefbtitle {Fluid Mechanics} {Fluid mechanics}\ (\PrintOrdinal{2}\ \BEd,
  \BVOL~6).
\newblock
\APACaddressPublisher{}{Elsevier}.
\PrintBackRefs{\CurrentBib}

\bibitem [\protect \citeauthoryear {%
Marshall%
\ \BBA {} Schott%
}{%
Marshall%
\ \BBA {} Schott%
}{%
{\protect \APACyear {1999}}%
}]{%
marshall1999open}
\APACinsertmetastar {%
marshall1999open}%
\begin{APACrefauthors}%
Marshall, J.%
\BCBT {}\ \BBA {} Schott, F.%
\end{APACrefauthors}%
\unskip\
\newblock
\APACrefYearMonthDay{1999}{}{}.
\newblock
{\BBOQ}\APACrefatitle {Open-ocean convection: Observations, theory, and models}
  {Open-ocean convection: Observations, theory, and models}.{\BBCQ}
\newblock
\APACjournalVolNumPages{Reviews of Geophysics}{37}{1}{1--64}.
\PrintBackRefs{\CurrentBib}

\bibitem [\protect \citeauthoryear {%
M{\"u}ller%
}{%
M{\"u}ller%
}{%
{\protect \APACyear {1995}}%
}]{%
muller1995ertel}
\APACinsertmetastar {%
muller1995ertel}%
\begin{APACrefauthors}%
M{\"u}ller, P.%
\end{APACrefauthors}%
\unskip\
\newblock
\APACrefYearMonthDay{1995}{}{}.
\newblock
{\BBOQ}\APACrefatitle {Ertel's potential vorticity theorem in physical
  oceanography} {Ertel's potential vorticity theorem in physical
  oceanography}.{\BBCQ}
\newblock
\APACjournalVolNumPages{Reviews of Geophysics}{33}{1}{67--97}.
\PrintBackRefs{\CurrentBib}

\bibitem [\protect \citeauthoryear {%
Pauluis%
}{%
Pauluis%
}{%
{\protect \APACyear {2008}}%
}]{%
p08}
\APACinsertmetastar {%
p08}%
\begin{APACrefauthors}%
Pauluis, O.%
\end{APACrefauthors}%
\unskip\
\newblock
\APACrefYearMonthDay{2008}{}{}.
\newblock
{\BBOQ}\APACrefatitle {Thermodynamic consistency of the anelastic approximation
  for a moist atmosphere} {Thermodynamic consistency of the anelastic
  approximation for a moist atmosphere}.{\BBCQ}
\newblock
\APACjournalVolNumPages{J. Atmos. Sci.}{65}{8}{2719--2729}.
\PrintBackRefs{\CurrentBib}

\bibitem [\protect \citeauthoryear {%
Peterson%
\ \protect \BOthers {.}}{%
Peterson%
\ \protect \BOthers {.}}{%
{\protect \APACyear {2006}}%
}]{%
peterson2006trajectory}
\APACinsertmetastar {%
peterson2006trajectory}%
\begin{APACrefauthors}%
Peterson, B\BPBI J.%
, McClelland, J.%
, Curry, R.%
, Holmes, R\BPBI M.%
, Walsh, J\BPBI E.%
\BCBL {}\ \BBA {} Aagaard, K.%
\end{APACrefauthors}%
\unskip\
\newblock
\APACrefYearMonthDay{2006}{}{}.
\newblock
{\BBOQ}\APACrefatitle {Trajectory shifts in the {A}rctic and subarctic
  freshwater cycle} {Trajectory shifts in the {A}rctic and subarctic freshwater
  cycle}.{\BBCQ}
\newblock
\APACjournalVolNumPages{Science}{313}{5790}{1061--1066}.
\PrintBackRefs{\CurrentBib}

\bibitem [\protect \citeauthoryear {%
Pollard%
\ \BBA {} Regier%
}{%
Pollard%
\ \BBA {} Regier%
}{%
{\protect \APACyear {1990}}%
}]{%
pollard1990large}
\APACinsertmetastar {%
pollard1990large}%
\begin{APACrefauthors}%
Pollard, R.%
\BCBT {}\ \BBA {} Regier, L.%
\end{APACrefauthors}%
\unskip\
\newblock
\APACrefYearMonthDay{1990}{}{}.
\newblock
{\BBOQ}\APACrefatitle {Large variations in potential vorticity at small spatial
  scales in the upper ocean} {Large variations in potential vorticity at small
  spatial scales in the upper ocean}.{\BBCQ}
\newblock
\APACjournalVolNumPages{Nature}{348}{6298}{227--229}.
\PrintBackRefs{\CurrentBib}

\bibitem [\protect \citeauthoryear {%
Rhines%
}{%
Rhines%
}{%
{\protect \APACyear {1986}}%
}]{%
rhines1986vorticity}
\APACinsertmetastar {%
rhines1986vorticity}%
\begin{APACrefauthors}%
Rhines, P\BPBI B.%
\end{APACrefauthors}%
\unskip\
\newblock
\APACrefYearMonthDay{1986}{}{}.
\newblock
{\BBOQ}\APACrefatitle {Vorticity dynamics of the oceanic general circulation}
  {Vorticity dynamics of the oceanic general circulation}.{\BBCQ}
\newblock
\APACjournalVolNumPages{Annual Review of Fluid Mechanics}{18}{1}{433--497}.
\PrintBackRefs{\CurrentBib}

\bibitem [\protect \citeauthoryear {%
Rossby%
}{%
Rossby%
}{%
{\protect \APACyear {1940}}%
}]{%
rossby1940planetary}
\APACinsertmetastar {%
rossby1940planetary}%
\begin{APACrefauthors}%
Rossby, C\BPBI G.%
\end{APACrefauthors}%
\unskip\
\newblock
\APACrefYearMonthDay{1940}{}{}.
\newblock
{\BBOQ}\APACrefatitle {Planetary flow patterns in the atmosphere} {Planetary
  flow patterns in the atmosphere}.{\BBCQ}
\newblock
\APACjournalVolNumPages{Quart. J. Roy. Met. Soc}{66 (Suppl.)}{}{68--87}.
\PrintBackRefs{\CurrentBib}

\bibitem [\protect \citeauthoryear {%
Salmon%
}{%
Salmon%
}{%
{\protect \APACyear {1998}}%
}]{%
salmon1998lectures}
\APACinsertmetastar {%
salmon1998lectures}%
\begin{APACrefauthors}%
Salmon, R.%
\end{APACrefauthors}%
\unskip\
\newblock
\APACrefYear{1998}.
\newblock
\APACrefbtitle {Lectures on {G}eophysical {F}luid {D}ynamics} {Lectures on
  {G}eophysical {F}luid {D}ynamics}.
\newblock
\APACaddressPublisher{}{Oxford University Press}.
\PrintBackRefs{\CurrentBib}

\bibitem [\protect \citeauthoryear {%
Schubert%
, Hausman%
, Garcia%
, Ooyama%
\BCBL {}\ \BBA {} Kuo%
}{%
Schubert%
\ \protect \BOthers {.}}{%
{\protect \APACyear {2001}}%
}]{%
schubert2001potential}
\APACinsertmetastar {%
schubert2001potential}%
\begin{APACrefauthors}%
Schubert, W\BPBI H.%
, Hausman, S\BPBI A.%
, Garcia, M.%
, Ooyama, K\BPBI V.%
\BCBL {}\ \BBA {} Kuo, H\BHBI C.%
\end{APACrefauthors}%
\unskip\
\newblock
\APACrefYearMonthDay{2001}{}{}.
\newblock
{\BBOQ}\APACrefatitle {Potential vorticity in a moist atmosphere} {Potential
  vorticity in a moist atmosphere}.{\BBCQ}
\newblock
\APACjournalVolNumPages{J. Atmos. Sci.}{58}{21}{3148--3157}.
\PrintBackRefs{\CurrentBib}

\bibitem [\protect \citeauthoryear {%
Smith%
\ \BBA {} Stechmann%
}{%
Smith%
\ \BBA {} Stechmann%
}{%
{\protect \APACyear {2017}}%
}]{%
smith2017precipitating}
\APACinsertmetastar {%
smith2017precipitating}%
\begin{APACrefauthors}%
Smith, L\BPBI M.%
\BCBT {}\ \BBA {} Stechmann, S\BPBI N.%
\end{APACrefauthors}%
\unskip\
\newblock
\APACrefYearMonthDay{2017}{}{}.
\newblock
{\BBOQ}\APACrefatitle {Precipitating quasigeostrophic equations and potential
  vorticity inversion with phase changes} {Precipitating quasigeostrophic
  equations and potential vorticity inversion with phase changes}.{\BBCQ}
\newblock
\APACjournalVolNumPages{J. Atmos. Sci.}{74}{10}{3285--3303}.
\PrintBackRefs{\CurrentBib}

\bibitem [\protect \citeauthoryear {%
Taylor%
\ \BBA {} Fournier%
}{%
Taylor%
\ \BBA {} Fournier%
}{%
{\protect \APACyear {2010}}%
}]{%
taylor2010compatible}
\APACinsertmetastar {%
taylor2010compatible}%
\begin{APACrefauthors}%
Taylor, M\BPBI A.%
\BCBT {}\ \BBA {} Fournier, A.%
\end{APACrefauthors}%
\unskip\
\newblock
\APACrefYearMonthDay{2010}{}{}.
\newblock
{\BBOQ}\APACrefatitle {A compatible and conservative spectral element method on
  unstructured grids} {A compatible and conservative spectral element method on
  unstructured grids}.{\BBCQ}
\newblock
\APACjournalVolNumPages{Journal of Computational Physics}{229}{17}{5879--5895}.
\PrintBackRefs{\CurrentBib}

\bibitem [\protect \citeauthoryear {%
Thomson%
}{%
Thomson%
}{%
{\protect \APACyear {1867}}%
}]{%
thomson1867}
\APACinsertmetastar {%
thomson1867}%
\begin{APACrefauthors}%
Thomson, W.%
\end{APACrefauthors}%
\unskip\
\newblock
\APACrefYearMonthDay{1867}{}{}.
\newblock
{\BBOQ}\APACrefatitle {4. On vortex atoms} {4. on vortex atoms}.{\BBCQ}
\newblock
\APACjournalVolNumPages{Proceedings of the Royal Society of
  Edinburgh}{6}{}{94-105}.
\newblock
\begin{APACrefDOI} \doi{10.1017/S0370164600045430} \end{APACrefDOI}
\PrintBackRefs{\CurrentBib}

\bibitem [\protect \citeauthoryear {%
Thorpe%
, Volkert%
\BCBL {}\ \BBA {} Ziemia{\'n}ski%
}{%
Thorpe%
\ \protect \BOthers {.}}{%
{\protect \APACyear {2003}}%
}]{%
thorpe2003bjerknes}
\APACinsertmetastar {%
thorpe2003bjerknes}%
\begin{APACrefauthors}%
Thorpe, A\BPBI J.%
, Volkert, H.%
\BCBL {}\ \BBA {} Ziemia{\'n}ski, M\BPBI J.%
\end{APACrefauthors}%
\unskip\
\newblock
\APACrefYearMonthDay{2003}{}{}.
\newblock
{\BBOQ}\APACrefatitle {The {B}jerknes' Circulation Theorem: A Historical
  Perspective: A Historical Perspective} {The {B}jerknes' circulation theorem:
  A historical perspective: A historical perspective}.{\BBCQ}
\newblock
\APACjournalVolNumPages{Bull. Am. Meteorol. Soc.}{84}{4}{471--480}.
\PrintBackRefs{\CurrentBib}

\bibitem [\protect \citeauthoryear {%
Thuburn%
}{%
Thuburn%
}{%
{\protect \APACyear {2008}}%
}]{%
thuburn2008some}
\APACinsertmetastar {%
thuburn2008some}%
\begin{APACrefauthors}%
Thuburn, J.%
\end{APACrefauthors}%
\unskip\
\newblock
\APACrefYearMonthDay{2008}{}{}.
\newblock
{\BBOQ}\APACrefatitle {Some conservation issues for the dynamical cores of
  {NWP} and climate models} {Some conservation issues for the dynamical cores
  of {NWP} and climate models}.{\BBCQ}
\newblock
\APACjournalVolNumPages{Journal of Computational Physics}{227}{7}{3715--3730}.
\PrintBackRefs{\CurrentBib}

\bibitem [\protect \citeauthoryear {%
Vallis%
}{%
Vallis%
}{%
{\protect \APACyear {2017}}%
}]{%
vallis2017atmospheric}
\APACinsertmetastar {%
vallis2017atmospheric}%
\begin{APACrefauthors}%
Vallis, G\BPBI K.%
\end{APACrefauthors}%
\unskip\
\newblock
\APACrefYear{2017}.
\newblock
\APACrefbtitle {Atmospheric and {O}ceanic {F}luid {D}ynamics} {Atmospheric and
  {O}ceanic {F}luid {D}ynamics}.
\newblock
\APACaddressPublisher{}{Cambridge University Press}.
\PrintBackRefs{\CurrentBib}

\bibitem [\protect \citeauthoryear {%
Webb%
\ \BBA {} Mace%
}{%
Webb%
\ \BBA {} Mace%
}{%
{\protect \APACyear {2015}}%
}]{%
webb2015potential}
\APACinsertmetastar {%
webb2015potential}%
\begin{APACrefauthors}%
Webb, G.%
\BCBT {}\ \BBA {} Mace, R.%
\end{APACrefauthors}%
\unskip\
\newblock
\APACrefYearMonthDay{2015}{}{}.
\newblock
{\BBOQ}\APACrefatitle {Potential vorticity in magnetohydrodynamics} {Potential
  vorticity in magnetohydrodynamics}.{\BBCQ}
\newblock
\APACjournalVolNumPages{Journal of Plasma Physics}{81}{1}{}.
\PrintBackRefs{\CurrentBib}

\bibitem [\protect \citeauthoryear {%
Wetzel%
, Smith%
, Stechmann%
, Martin%
\BCBL {}\ \BBA {} Zhang%
}{%
Wetzel%
\ \protect \BOthers {.}}{%
{\protect \APACyear {2020}}%
}]{%
wetzel2020potential}
\APACinsertmetastar {%
wetzel2020potential}%
\begin{APACrefauthors}%
Wetzel, A\BPBI N.%
, Smith, L\BPBI M.%
, Stechmann, S\BPBI N.%
, Martin, J\BPBI E.%
\BCBL {}\ \BBA {} Zhang, Y.%
\end{APACrefauthors}%
\unskip\
\newblock
\APACrefYearMonthDay{2020}{}{}.
\newblock
{\BBOQ}\APACrefatitle {Potential Vorticity and Balanced and Unbalanced
  Moisture} {Potential vorticity and balanced and unbalanced moisture}.{\BBCQ}
\newblock
\APACjournalVolNumPages{J. Atmos. Sci.}{77}{6}{1913--1931}.
\PrintBackRefs{\CurrentBib}

\end{thebibliography}

%
%




%
%
%
%
%

\end{document}